# Polarized light emission from graphene induced by terahertz pulses

*I. V. Oladyshkin\*, S. B. Bodrov, A. V. Korzhimanov, A. A. Murzanev, Yu. A. Sergeev, A. I. Korytin, M. D. Tokman and A. N. Stepanov*

*\*oladyshkin@ipfran.ru*

*Institute of Applied Physics of the Russian Academy of Sciences, 603950, Nizhny Novgorod, Russia*

Spontaneous optical emission of graphene irradiated by intense single-cycle terahertz pulses was investigated experimentally and explained theoretically. We found that emitted photons are polarized predominantly perpendicular to the electric field of the terahertz pulse, which proves that the terahertz field not only heats the electrons, but also creates a strongly nonequilibrium momentum distribution. Comparison of the measured optical spectrum and polarization anisotropy with the results of numerical modeling allowed us to estimate a momentum isotropization time for electrons in graphene to be ~25 fs and roughly reconstruct the distribution function evolution in *k*-space.

Graphene is one of perspective materials attracting attention due to unusual properties of charge carriers and unique characteristics like 2D geometry, high nonlinearity, record electrical and thermal conductivity etc. Nonlinear and quantum optics of graphene is now actively investigated [1-14]. Over the past few years, ultrafast kinetics of electrons in graphene became one of the central topics in a large number of fundamental and applied studies of various subpicosecond processes. In particular, it was found that femto- and picosecond relaxation processes play a key role in such effects as nonlinear terahertz response [7, 8, 15], surface plasmon generation [10, 11, 16, 17], nonlinear transmission [8, 12], high harmonic generation [18-24], inverse Faraday effect and generation of edge photocurrents [9, 25] etc.

At the same time, dynamics of elementary relaxation processes at ultrashort timescales in graphene and other materials is still a relevant problem due to limitations of available experimental techniques and high complexity of first-principle modeling. Here we should mention a recent attempt to control and measure *e-e* electron scattering length in graphene in a state-of-art experiment performed by group of M. Polini and A. K. Geim [26]. To date, various pump-probe experiments (terahertz pump – optical probe [27], optical pump – terahertz probe [28], optical pump – optical probe [29-31]) and theoretical works [32–35] were devoted to the investigation of ultrafast carrier dynamics in graphene, especially the thermalization and recombination processes. In particular, it was demonstrated that the characteristic time of interband recombination (~1 ps) significantly exceeds the times of carrier thermalization and cooling (100–300 fs) [12, 27].

Recently, in Ref. [37] terahertz-field-induced spontaneous optical emission in the range of 350–600 nm was observed from monolayer graphene on glass substrate (at peak terahertz electric fields from 100 to 250 kV/cm). To explain the experimental results an analytical theory based on electron-hole pair production by Schwinger-like (or Landau-Zener) transitions in strong electric field was proposed. The thermal mechanism which can also contribute to the optical emission was estimated to be two orders weaker than the measured one. At the same time, in Ref. [27] the dominant role of impact ionization, or interband reverse Auger recombination, in the process of electron-hole pair production by strong terahertz field was stated[1].

---

[1] Interband reverse Auger recombination is forbidden in ideal 2D Dirac systems due to the energy and momentum conservation laws [38–40]. In general case the efficiency of interband Auger processes is determined by the lattice imperfection (e.g. trigonal warping [41]) or by the distortion of conservation laws for two-particle collisions (which was a fitting parameter in [27]) in the presence of an additional interaction – with phonons [39, 40] or others.

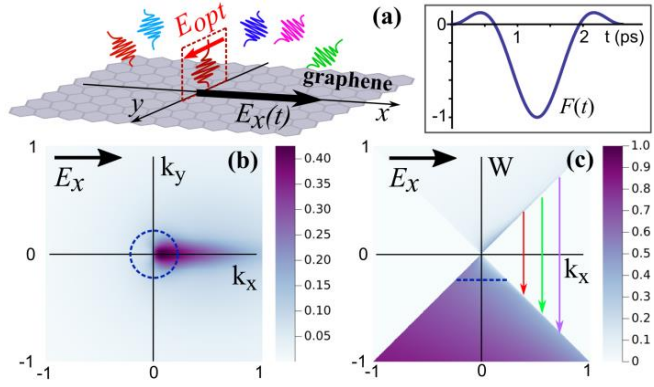

Fig. 1. (a) Geometry of the problem: terahertz electric filed $E_x(t) = E_{THz} \cdot F(t)$ oriented along *x*-axis induces spontaneous optical photons of different wavelengths with the dominant polarization of electric field $E_{opt}$ along *y*-axis. Inset: normalized waveform of the THz pulse $F(t)$ used in the numerical modelling. (b) Numerically calculated distribution of electrons in *k*-space of the conduction band at the moment of peak terahertz field ($E_{THz} = 190$ kV/cm); a dashed blue circle depicts the initial Fermi level in the valence band. (c) The same distribution in energy space; the initial Fermi level is shown by a dashed blue line, possible direct interband transitions are shown by color arrows. In (b) and (c) color bars represent occupation probability and the unit scales of vertical and horizontal axes correspond to $W = 1\ eV$ and $\hbar k_{x,y} v_F = 1\ eV$.

In this Letter we provide more clear experimental evidences of strongly non-thermal character of terahertz-field-induced optical emission from graphene. For this we measured spectral and polarization properties of the optical emission in the spectral range of 350-1050 nm. In combination with the numerical modeling, this allows us to prove distribution function in *k*-space to be significantly anisotropic and to estimate a collision frequency and a temperature of excited electrons.

Let us start with a theoretical model. Schematically, the optical emission from graphene induced by strong terahertz field can be presented as a four-step process (Fig. 1):
- strong electron acceleration by the electric field of the terahertz pulse $E_x(t) = E_{THz} \cdot F(t)$, where $E_{THz}$ and $F(t)$ are peak value and normalized waveform of terahertz field, respectively;
- electron-hole pair production by Schwinger-like (or Landau-Zener) transitions, leading to local population inversion in *k*-space;
- intraband isotropization of electrons and holes (both in the conduction and valence bands);
- spontaneous interband electron-hole recombination accompanied by the optical photon emission.

In our initial study [37] the intraband isotropization was assumed instantaneous. Here we do not use such assumption and model dynamics of the electronic distribution function in *k*-space including both the isotropization and recombination. In the calculations we use the density matrix equations for a two-band system in a homogeneous electric field [23] which is a version of the semiconductor Bloch equations:

$$\left(\frac{\partial}{\partial t} + i\omega_k + \gamma_a - \frac{eE_x(t)}{\hbar}\frac{\partial}{\partial k_x}\right)\rho_k = -i\frac{\Omega_k}{2}[n_c(\mathbf{k}) - n_v(\mathbf{k})], \quad (1)$$

$$\frac{\partial n_{c,v}(\mathbf{k})}{\partial t} - \frac{e}{\hbar}E_x(t)\frac{\partial n_{c,v}(\mathbf{k})}{\partial k_x} = \mp i\Omega_k \operatorname{Im}(\rho_\mathbf{k}) + R_{c,v}(\mathbf{k}), \quad (2)$$

where $e$ is the elementary charge, **k** is the electron wavevector, $n_{c,v}(\mathbf{k})$ are the electron populations in the conduction and valence bands respectively, i.e. the diagonal terms of 2×2 density matrix, $\rho_\mathbf{k} = \rho_{cv;\mathbf{kk}}$ is the interband quantum coherence (off-diagonal terms), $\omega_k = 2v_F k$ is a local interband transition frequency at a given point of the *k*-space, $v_F$ is the Fermi velocity (~$10^8$ cm/s), $\Omega_k = \frac{e}{k\hbar}\sin\theta_\mathbf{k} E_x$ is a local Rabi frequency, $\theta_\mathbf{k}$ is an angle between **k** and *x*-axis, $\gamma_a^{-1}$ is the relaxation time for the interband quantum coherence and $R_{c,v}(\boldsymbol{k})$ are phenomenological operators of the population relaxation in the conduction and valence bands respectively.

The operators of population relaxation are set in the simplest form (BGK, Bhatnagar–Gross–Krook) with two characteristic timescales: "fast" intraband relaxation and "slow" interband recombination (described by $\gamma_r$):

$$R_{c,v}(\boldsymbol{k}) = -\gamma_a[n_{c,v}(\boldsymbol{k}) - n^F_{c,v}(k)] - \gamma_r[n_{c,v}(\boldsymbol{k}) - n^F_{0(c,v)}(k)]. \quad (3)$$

The intraband relaxation constant[1] is supposed to be equal to the coherence relaxation rate $\gamma_a$ in our calculation.

Here

$$n^F_{0(c,v)}(k) = \frac{1}{1+\exp[(\pm v_F \hbar k/2 - \mu_0)/T]} \quad (4)$$

is a globally equilibrium Fermi distribution with the initial chemical potential $\mu_0$, and

$$n^F_{c,v}(k) = \frac{1}{1+\exp[(\pm v_F \hbar k/2 - \mu_{c,v})/T]} \quad (5)$$

are quasiequilibrium local Fermi distributions in the conduction (*c*) and valence bands (*v*) respectively, having a preset temperature of electrons $T$ and time-dependent chemical potentials $\mu_{c,v}$. The chemical potentials $\mu_{c,v}$ are calculated independently using the normalization relations which include the numbers of charge carriers in each band in the current moment of time:

$$\iint_\infty n_c(\boldsymbol{k})d^2k = 2\pi\int_0^\infty n^F_c(k)kdk, \quad \iint_\infty [1 - n_v(\boldsymbol{k})]d^2k = 2\pi\int_0^\infty [1 - n^F_v(k)]kdk, \quad (6)$$

Basing on previously reported results, we suppose that in our experimental conditions we operate slightly below the region of temperature saturation caused by the strong heat absorption by optical phonons with the excitation energy of about 0.2 eV [37, 43, 44]. Because of that involving the electron temperature dependence on the peak terahertz field is principle for the numerical model developing. For simplicity we assume linear dependence $T(E_{THz})$. As for the interband recombination rate, we found that variation of $\gamma_r^{-1}$ in the range of 0.3–1 ps didn't significantly affect the simulations, so, we put it to be equal to 500 fs. The intraband relaxation time $\gamma_a^{-1}$, in contrast, significantly alters the polarization properties of the spontaneous emission, which allow us to estimate the anisotropy relaxation rate from experimental data.

The numbers of photons with *x*- and *y*-polarizations, spontaneously emitted due to the interband transitions, were calculated using the analytical relations given in the Appendix of Ref. [37] (where the probability of spontaneous emission of photons with *x*- and *y*-polarization in the graphene with an arbitrary electron distribution $n_{c,v}(\boldsymbol{k})$ is found in general form). Note that at room temperature the interband relaxation is caused mostly by electron-phonon interaction, so the spontaneous emission of photons is a process with a relatively low probability which modifies the distribution function insignificantly. Technical details of the numerical scheme can be found in Supplementary Materials.

---

[1] The discussed phenomenological model of relaxation might, in principle, include two different characteristic timescales of intraband relaxation: the isotropization time $\gamma_{a1}^{-1}$ and several times greater thermalization time $\gamma_{a2}^{-1}$ instead of the universal constant $\gamma_a^{-1}$. However, the recombination time $\gamma_r^{-1}$ undoubtedly exceeds both the $\gamma_{a1}^{-1}$ and $\gamma_{a2}^{-1}$, so the quasiequilibrium distributions inside each band are formed faster than the totally equilibrium distribution. That is why this kind of model complication cannot significantly influence the experimental data interpretation, but, on the contrary, adds another one poorly known free parameter.

A typical calculated distribution function of electrons in a strong terahertz field is shown in Fig. 1 (b, c). The region of significant population perturbation extends up to the energy level of 0.5–1.0 eV, which corresponds to direct interband transitions with 1–2 eV energy. The calculations predict an intense electron-hole pair production with the density $N_c \cong 2.5 \cdot 10^{12}\ cm^{-2}$ and significant anisotropy of the electron momenta. This density can be also calculated analytically in the ballistic ionization model[1] [37]:

$$\frac{dN_c}{dt} = \left(\frac{eE_x}{\hbar}\right)^{\frac{3}{2}} \frac{(\bar{n}_v - \bar{n}_c)}{\pi^2 \sqrt{v_F}}, \tag{7}$$

where $\bar{n}_{v,c}$ are averaged populations in the valence and conduction bands respectively. Equation (7) predicts $N_c \cong 5 \cdot 10^{12}\ cm^{-2}$ for the same parameters as used in Fig. 1, which overestimates the density obtained in the numerical modelling because Eq. (7) does not take into account the interband recombination.

Distribution function anisotropy can be also estimated analytically in the simplest model of relaxation. For this we introduce a parameter describing a relative deformation of the distribution function $\eta = \delta k_x / k_F$, where $k_F = \sqrt{\pi N_c}$ is a "local" Fermi wavenumber in the conduction band and $\delta k_x$ is the stationary shift of the distribution function in $k$-space. In the Drude model $\delta k_x$ can be calculated as

$$\delta k_x = \frac{eE_x}{\hbar \gamma_a}. \tag{8}$$

Using Eq. (7) in assumption that initially $\bar{n}_v \approx 1$, $\bar{n}_c \ll 1$ and introducing character duration of the terahertz pulse $\tau_{THz}$, we arrive to:

$$\eta = \frac{\pi}{\gamma_a} \left(\frac{eE_{THz} v_F}{\hbar \tau_{THz}^2}\right)^{\frac{1}{4}}. \tag{9}$$

Thus, the relative deformation of the distribution function is proportional to $E_{THz}^{1/4}$, so the dependence is rather weak. This is a consequence of the competition between two physical effects: the growth of the local Fermi level in the conduction band and the increase of the distribution function shift. At the same time, the deformation parameter $\eta$ determines the ratio between the numbers of spontaneous photons with $y$- and $x$-polarizations, which follows from the expressions for the probability of spontaneous emission of an electron with given momentum direction (see Appendix in Ref. [37]).

An experimental setup used to investigate the polarization and spectral properties of terahertz-field-induced luminescence of graphene is shown in Fig. 2 and in more detail described elsewhere [37, 45]. Single-cycle terahertz pulses generated by tilted-pulse-front technique in a LiNbO$_3$ crystal were strongly focused on a graphene sample. The terahertz electric field was vertically polarized (along $x$-axis) and has peak magnitude up to ~250 kV/cm. The optical luminescence emitted in forward direction was collected by F1 lens (D = 45 mm and NA = 1) and focused by an objective lens F2 (Canon EF 50 mm F/1.4) on a cooled CCD camera (SPEC-10:400BR Princeton Instruments). Camera exposure was set to 100 seconds. A wide polarizer (25x25 mm, Thorlabs WP25L-UB) and a set of calibrated color filters were placed after F1 lens to measure the polarization and spectral properties of the luminescence. The number of emitted photons was calculated as a sum of CCD counts over an illuminated region (covering typically 50x50 pixels) with account of transparency of the corresponding color filters and spectral responsivity of the detection system calibrated with a blackbody source (LS-1-LL) with the temperature of 2800 K.

---

[1] Eq. (7) is valid when the time of Landau-Zener transition $\delta t = \sqrt{\hbar / eE_x v_F}$ is less than the momentum relaxation time $\gamma_a^{-1}$ [37]. Typically, this condition is fulfilled for THz fields stronger than 70-100 kV/cm ($\delta t = 8 - 10\ fs$), which is also a threshold of the intense pair production.

We used two samples: monolayer graphene on polyethylene terephthalate (PET) and glass substrates. Note that a doping of graphene is of p-type in both cases and the characteristic Fermi levels are –0.2 eV for glass and –0.35 eV for PET [42] substrates. Due to Fresnel reflection, the terahertz electric field in graphene was reduced by a factor of 1.25 and 1.67 for PET and glass substrates, respectively.

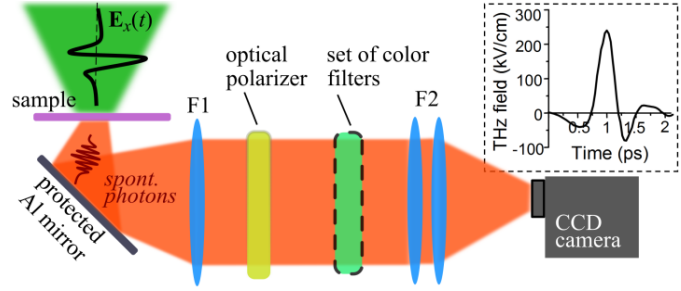

Fig. 2. Experimental setup. Inset: waveform of the terahertz pulse measured by time-domain spectroscopy technique.

Figure 3 shows a number of emitted photons $N_{ph}$ as a function of the peak terahertz electric field $E_{THz}$ in three selected spectral ranges: 400-600 nm, 600-850 nm, and 850-1100 nm (see inset in Fig. 3). Similar to previous results [37, 45], the luminescence in each spectral range increases rapidly with the field strength. The faster gradient is observed for the shorter wavelength. From the numerical calculations we found that $N_{ph}(E_{THz})$ is highly sensitive to the temperature $T$. Satisfactory coincidence with the experiment was obtained for a linear fitting of the phenomenological dependence $T(E_{THz})$ introduced in the theoretical model:

$$T(E_{THz}) = 44\ meV + 1.1 meV \cdot E_{THz}[kV/cm]. \tag{10}$$

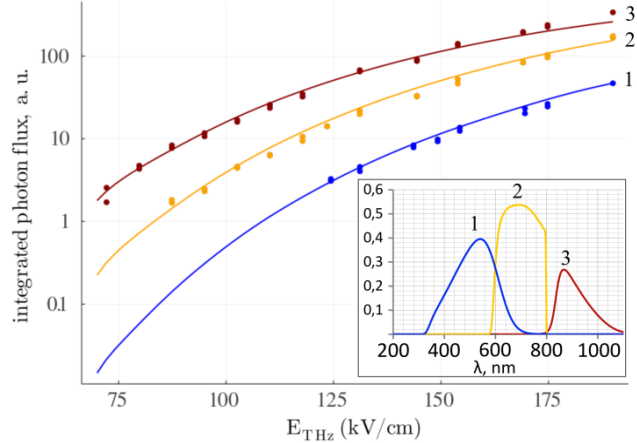

Fig. 3. Number of terahertz-field-induced optical photons in different spectral ranges as a function of peak terahertz field for the graphene on PET substrate. Experimental data and numerical calculations are represented by circles and color lines, respectively. Spectral characteristics of the used color filters are shown in the inset; the experimental points for the photon flux are plotted after normalization on the integral transitivity of the corresponding filter. The maximal absolute number of detected photons for the filter #3 is about $1 \cdot 10^6$ per $6 \cdot 10^4$ shots at 190 kV/cm.

From the experiment we found that the polarization anisotropy parameter σ, which is a ratio of photon numbers detected after the horizontal and vertical polarizers, remained roughly constant σ ≈ 1.35 in the available range of terahertz field magnitudes 120–190 kV/cm (see Fig. 4). Numerical modelling showed that the dependence σ($E_{THz}$) significantly modifies with the change of anisotropy relaxation time $\gamma_a^{-1}$. In particular, at $\gamma_a^{-1} = 45\ fs$ the polarization ratio decreases from 2.1 to 1.5 with the terahertz field increase from 120 to 190 kV/cm, while at $\gamma_a^{-1} = 15\ fs$, in contrast, this dependence occurs to be slightly growing near the value σ ≈ 1.1. The best coincidence with the experimental data (the absolute value and the near-zero slope) is achieved for $\gamma_a^{-1} = 26 \pm 5\ fs$. Note that the experimental results qualitatively agree with the analytical Eq. (9), predicting slowly growing dependence σ($E_{THz}$) and giving the deformation parameter $\eta \cong 1.5$ for $\tau_{THz} = 200\ fs$ (see inset in Fig. 2) and $\gamma_a^{-1} = 26\ fs$, which corresponds well to the measured σ.

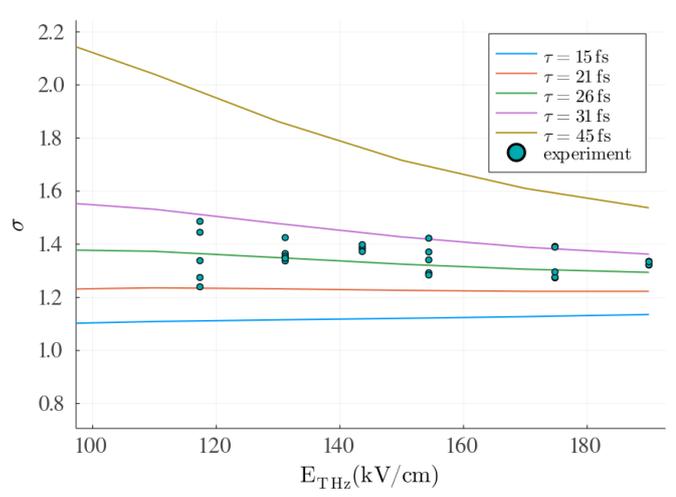

Fig. 4. Polarization anisotropy σ as a function of the peak terahertz field magnitude. Circles – experimental data, color lines – numerical modeling results for different relaxation times $\gamma_a^{-1}$ specified in the inset.

The optical emission spectra from graphene on the PET and glass substrates (measured for the incident terahertz field of 250 kV/cm) are shown in Fig. 5. For the sample with PET substrate the emission is about two times more intensive than for the sample with glass substrate. This is a result of different terahertz pulse magnitudes in graphene on PET and glass substrates (190 kV/cm and 140 kV/cm, respectively) which gives, according Eq. (10), different effective electron temperatures (0.25 eV and 0.2 eV respectively). Numerical calculations for these samples without any adjustable parameters, give emission spectra which agree well with the experiment both in the shape and the magnitude (see Fig.5). This fact corroborates the proposed numerical model.

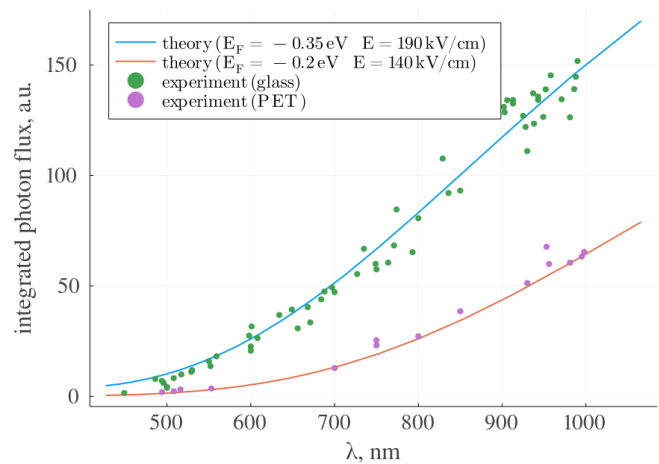

Fig. 5. Terahertz-field-induced luminescence spectra from graphene on PET (green points and blue curve) and glass (violet points and red curve) substrates. Experimental data – points, numerical modeling – curves. Initial Femi levels and the magnitude of the terahertz field $E_{THz}$ used in numerical modeling are indicated in the inset.

To conclude, we investigated spontaneous optical emission of graphene with strongly nonequilibrium charge carrier distribution excited by a single-cycle terahertz pulse. This broadband optical radiation has a quasithermal spectrum (in the range of 350-1050 nm) but occurs to be significantly polarized with prevailing polarization perpendicularly to the incident terahertz field. The polarization ratio was found to be approximately constant (about 1.35) for the terahertz field magnitudes from 120 to 190 kV/cm. We interpreted these facts as a consequence of a strong deformation of the electronic distribution function in graphene, which remains almost constant due to the competition between the field acceleration, scattering and temperature growth processes.

Comparison of the experimental data with the numerical calculations allowed us to find the intraband relaxation time 26±5 fs and to reconstruct the electronic temperature dependence on the terahertz field. The best fitting was achieved for $T(E_{THz}) = 44\ meV + 1.1 meV \cdot E_{THz}[kV/cm]$ which can be used for further validation of microscopic heating models.

Finally, we would like to emphasize that the observation of a spontaneous optical emission after a sub-picosecond terahertz excitation is a powerful experimental method for obtaining new data on ultrafast kinetics of nonequilibrium electrons in graphene, as well as in other materials. Being polarization-sensitive by its nature, this method gives a possibility to study an ultrafast isotropization stage of a distribution function evolution.

**Acknowledgements**. The experimental part of the work was financed by RFBR (grant #18-29-19091_mk); operation of the laser system was supported by the Ministry of Science and Higher Education of the Russian Federation, state assignment for the Institute of Applied Physics RAS, project 0030-2021-0012. Personally I.V.O. thanks the Russian Science Foundation which supported the theoretical analysis and data interpretation (project #21-72-00076). The numerical simulations were performed on resources provided by the Joint Supercomputer Center of the Russian Academy of Sciences. The authors also thank prof. Alexey Belyanin for fruitful discussions.

**References**

1. F. Bonaccorso, Z. Sun, T. Hasan, and A. Ferrari, Nat. Photonics 4, 611 (2010).

2. T. Gu, N. Petrone, J. F. McMillan, A. van der Zande, M. Yu, G.-Q. Lo, D.-L. Kwong, J. Hone, and C. W. Wong, Nat. Photonics 6, 554 (2012).

3. D. Sun, C. Divin, J. Rioux, J. E. Sipe, C. Berger, W. A. de Heer, P. N. First, and T. B. Norris, Nano Lett. 10, 1293 (2010).

4. Y. Q. An, F. Nelson, J. U. Lee, and A. C. Diebold, Nano Lett. 13, 2104 (2013).

5. H. K. Avetissian, A. K. Avetissian, G. F. Mkrtchian, and K. V. Sedrakian, J. Nanophotonics 6, 061702 (2012).

6. T. Jiang, V. Kravtsov, M. Tokman, A. Belyanin, & M. B. Raschke, Nature Nanotechnology 14, 838-843 (2019).

7. H.A. Hafez, S. Kovalev, K.J. Tielrooij, M. Bonn, M. Gensch, D. Turchinovich, Advanced Optical Materials 8 (3), 1900771 (2020).

8. Hafez, H.A., Kovalev, S., Deinert, J.C. et al., Nature 561, 507–511 (2018).

9. M. Glazov and S. Ganichev, Phys. Rep. 535, 101 (2014).

10. X. Yao, M. Tokman, and A. Belyanin, Phys. Rev. Lett. 112, 055501 (2014).

11. T. J. Constant, S. M. Hornett, D. E. Chang, and E. Hendry, Nat. Phys. 12, 124 (2016).

12. J. C. König-Otto, M. Mittendorff, T. Winzer, F. Kadi, E. Malic, A. Knorr, C. Berger, W. A. de Heer, A. Pashkin, H. Schneider, M. Helm, and S. Winnerl, Phys. Rev. Lett. 117, 087401 (2016).

13. T. Li, L. Luo, M. Hupalo, J. Zhang, M. C. Tringides, J. Schmalian, and J. Wang, Phys. Rev. Lett. 108, 167401 (2012).

14. S. Wu, L. Mao, A. M. Jones, W. Yao, C. Zhang, and X. Xu, Nano Lett. 12, 2032 (2012)

15. C.B. Mendl, M. Polini, A. Lucas, Applied Physics Letters 118 (1), 013105 (2021)

16. D. A. Bandurin, D. Svintsov, I. Gayduchenko, S. G. Xu, A. Principi, M. Moskotin, I. Tretyakov, D. Yagodkin, S. Zhukov, T. Taniguchi, K. Watanabe, I. V. Grigorieva, M. Polini, G. N. Goltsman, A. K. Geim and G. Fedorov, Nature Communications 9 (1), 1-8 (2018)

17. Mikhail Tokman, Yongrui Wang, Ivan Oladyshkin, A. Ryan Kutayiah, and Alexey Belyanin, Phys. Rev. B 93, 235422 (2016)

18. S. A. Mikhailov, Physical Review B 103 (24), 245406 (2021)

19. S. A. Mikhailov, Physical Review B 100 (11), 115416 (2019)

20. J. Cheng, N. Vermeulen, and J. Sipe, Sci. Rep. 7, 43843 (2017).

21. J. J. Dean and H. M. van Driel, Appl. Phys. Lett. 95, 261910 (2009).

22. J. J. Dean and H. M. van Driel, Phys. Rev. B 82, 125411 (2010).

23. Mikhail Tokman, Sergei B. Bodrov, Yuri A. Sergeev, Alexei I. Korytin, Ivan Oladyshkin, Yongrui Wang, Alexey Belyanin, and Andrei N. Stepanov, Phys. Rev. B 99, 155411 (2019);

24. Yongrui Wang, Mikhail Tokman, and Alexey Belyanin, Phys. Rev. B 94, 195442 (2016)

25. I. D. Tokman, Q. Chen, I. A. Shereshevsky, V. I. Pozdnyakova, I. Oladyshkin, M. Tokman, A. Belyanin, Physical Review B 101 (17), 174429 (2020)

26. Kim, M., Xu, S.G., Berdyugin, A.I. et al., Nat Commun 11, 2339 (2020).

27. S. Tani, F. Blanchard, and K. Tanaka, Phys. Rev. Lett. 109, 166603 (2012).

28. K. J. Tielrooij, J. C. W. Song, S. A. Jensen, A. Centeno, A. Pesquera, A. Zurutuza Elorza, M. Bonn, L. S. Levitov, and F. H. L. Koppens, Nat. Phys. 9, 248 (2013).

29. M. Mittendorff, T. Winzer, E. Malic, A. Knorr, C. Berger, W. A. de Heer, H. Schneider, M. Helm, and S. Winnerl, Nano Lett. 14, 1504 (2014).

30. D. Brida, A. Tomadin, C. Manzoni, Y. J. Kim, A. Lombardo, S. Milana, R. R. Nair, K. S. Novoselov, A. C. Ferrari, G. Cerullo, and M. Polini, Nat. Commun. 4, 1987 (2013).

31. T. Ploetzing, T. Winzer, E. Malic, D. Neumaier, A. Knorr, and H. Kurz, Nano Lett. 14, 5371 (2014).

32. T. Winzer, E. Malic, and A. Knorr, Phys. Rev. B 87, 165413 (2013).

33. T. Winzer, A. Knorr, and E. Malic, Nano Lett. 10, 4839 (2010).

34. E. Malic, T. Winzer, E. Bobkin, and A. Knorr, Phys. Rev. B 84, 205406 (2011)

35. E. Malic, T. Winzer, and A. Knorr, Appl. Phys. Lett. 101, 213110 (2012).

36. G. Kane, M. Lazzeriand, and F. Mauri, J. Phys. Condens. Matter. 27, 164205 (2015).

37. I.V. Oladyshkin, S.B. Bodrov, Yu.A. Sergeev, A.I.Korytin, M.D Tokman, A.N. Stepanov, Phys. Rev. B 96, 155401 (2017).

38. M. S. Foster and I. L. Aleiner, Phys. Rev. B 79, 085415 (2009).

39. Justin C. W. Song, Klaas J. Tielrooij, Frank H. L. Koppens, and Leonid S. Levitov, Phys. Rev. B 87, 155429 (2013).

40. Brida, D. A. Tomadin, C. Manzoni, Y.J. Kim, A. Lombardo, S. Milana, R.R. Nair, K.S. Novoselov, A.C. Ferrari, G. Cerullo & M. Polini, Nat. Commun. 4:1987 (2013).

41. V. N. Kotov, B. Uchoa, V. M. Pereira, F. Guinea, & A. H. Castro Neto, Rev. Mod. Phys. 84, 1067–1125 (2012).

42. P.R. Whelan, Q. Shen, B. Zhou et al., 2D Materials 7, 035009 (2020).

43. C. H. Lui, K. F. Mak, J. Shan, and T. F. Heinz, Phys. Rev. Lett. 105, 127404 (2010).

44. H. Wang, J. H. Strait, P. A. George, S. Shivaraman, V. B. Shields, M. Chandrashekhar, J. Hwang, F. Rana, M. G. Spencer, C. S. Ruiz-Vargas, and J. Park, Appl. Phys. Lett. 96, 081917 (2010).

45. S. Bodrov, A. Murzanev, A. Korytin, and A. Stepanov, Opt. Lett. 46, 5946-5949 (2021).

**Supplementary Materials. Numerical modelling details**

The density matrix equations were discretized on a 2D rectangular grid and numerically integrated by an operator splitting method [A1] in the following way. An advection term $\partial/\partial t - eE/\hbar \partial/\partial k_{x,y}$ has been integrated by means of the positivity flux conservative scheme [A2] from Vasilek.jl project [A3]. Rabi terms and damping terms were calculated by the simple Euler scheme. A written code has been tested against a number of simple problems with known results and has been shown to be stable and reasonably accurate. In all simulations a $300 \times 100$ box has been used corresponding to physical dimensions of $[-1.5, 1.5] \times [-1.5, 1.5]$ measured in $\text{eV}/\hbar v_F$. A time step was equal to $0.4 \text{eV}^{-1} \cdot \hbar \approx 1.6$ fs and a total simulation time was equal to 7000 time steps or about 11.6 ps. Initially the Fermi distribution was set with a given Fermi energy of $-0.35\ eV$ for a PET substrate and $-0.2\ eV$ for a glass substrate. Other parameters were varied as indicated in the text.

[A1] *C.Z. Cheng and G. Knorr, The integration of the Vlasov equation in configuration space, J. Comput. Phys., 22 (1976), pp. 330–351*

[A2] *Filbet, F., Sonnendrücker, E. & Bertrand, P. Conservative Numerical Schemes for the Vlasov Equation. J. Comput. Phys. 172, 166–187 (2001)*

[A3] https://github.com/korzhimanov/Vasilek.jl